\documentclass[%
 reprint,
superscriptaddress,
longbibliography,
aip,
 amsmath,amssymb,
]{revtex4-1}
\setcitestyle{numbers,square}

\usepackage{braket}
\usepackage{graphicx}
\usepackage{dcolumn}
\usepackage{bm}
\usepackage[normalem]{ulem} 
\usepackage{color}

\usepackage{hyperref}
\hypersetup{
    colorlinks,%
    citecolor=blue,%
    linkcolor=blue,%
    urlcolor=blue
}

\newcommand{\orcid}[1]{\href{https://orcid.org/#1}{\includegraphics[width=8pt]{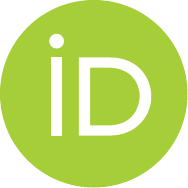}}}

\begin{document}

\title{Uncovering the Structural Evolution Arsenene on SiC Substrate}

\author{A. K. Okazaki \orcid{0000-0003-1935-9841}} 
\affiliation{Brazilian Nanotechnology National Laboratory, CNPEM, Campinas, SP, Brazil}

\author{R. Furlan de Oliveira \orcid{0000-0001-8980-3587}} 
\affiliation{Brazilian Nanotechnology National Laboratory, CNPEM, Campinas, SP, Brazil}

\author{R. L. H. Freire\orcid{0000-0002-4738-3120}} 
\affiliation{Ilum School of Science, CNPEM, Campinas, SP, Brazil}

\author{A. Fazzio\orcid{0000-0001-5384-7676}}
\email{adalberto.fazzio@ilum.cnpem.br}
\affiliation{Ilum School of Science, CNPEM, Campinas, SP, Brazil}
\affiliation{Center for Natural and Human Sciences, Federal University of ABC, Santo André, SP, Brazil}

\author{F. Crasto de Lima\orcid{0000-0002-2937-2620}}
\email{felipe.lima@ilum.cnpem.br}
\affiliation{Ilum School of Science, CNPEM, Campinas, SP, Brazil}

\date{\today}

\begin{abstract}

{Two-dimensional arsenic allotropes have been grown on metallic surfaces, while topological properties have been theoretically described on strained structures. Here we experimentally grow arsenene by molecular beam epitaxy over the insulating SiC substrate. The arsenene presents a flat structure with a strain field that follows the SiC surface periodicity. Our ab initio simulations, based on the density functional theory, corroborate the experimental observation. The strained structure presents a new arsenene allotrope with a triangular structure, rather than the honeycomb previously predicted for other pnictogens. This strained structure presents a Peierls-like transition leading to an indirect gap semiconducting behavior.}

\end{abstract}

\maketitle

\section{Introduction}

Strain engineering has become a powerful method to tailor atomically-thin bidimensional materials (2DMs) electronic and optical properties \cite{NATCOMMahn2017, ADVMATdai2019, ADVMATkim2022}. Theoretical and experimental studies have demonstrated a variety of phenomena related to the strain engineering of different 2DMs that range from the tuning of existing properties, e.g., the modulation of photoluminescence in monolayer MoS2 \cite{NLconley2013, ACSPHOTchowdhury2021, ACSMATlin2022}, up to the appearance of new or exotic effects, such as pseudo quantum Hall effect in graphene \cite{LSApeng2020, NATPHYSguinea2010}. All these efforts have sparked a tremendous rush for novel 2DMs exhibiting unique or enhanced properties that can be controlled by the application of an appropriate mechanical strain \cite{NATCOMMahn2017, ADVMATdai2019, ADVMATkim2022}. In seeking novel 2DMs that exhibit properties enabled by strain engineering, the group VA (pnictogens) has gained tremendous attention lately \cite{SMALLhu2022, CSRzhang2018, NLcosta2019, 2DMfocassio2021}. In particular, monoelemental arsenic layers (arsenene) have been predicted to exhibit a phase transition from normal to topological insulator upon appropriate strain \cite{PRBkamal2015}. Arsenene exhibits two allotropes, namely buckled (gray arsenic phase) and puckered (black arsenic phase) layers \cite{SMALLhu2022}. Puckered and buckled arsenene are indirect bandgap semiconductors, with $0.831$ and $1.635$\,eV bandgap, respectively \cite{PRBkamal2015}. Small strains have been reported to lead a transition to a direct bandgap semiconductor in puckered arsenene \cite{PRBkamal2015, PRBkamal2015}, with tunable bandgap energy and effective mass upon strain variation \cite{AIPADVwang2016}. Previous density functional theory (DFT) calculations show that tensile strains exceeding $11.7\%$ can make both arsenene polytypes behave as quantum spin hall (QSH) insulators \cite{NSzhang2015}. Larger strains $10-12\%$ or $>27\%$, have also been claimed to allow a semiconductor-to-metallic transition in both puckered and buckled arsenene \cite{AIPADVwang2016}. Such a large flexibility to manipulate different states and related properties of arsenene by means of strain can enable the development of a multitude of applications, including new-concept devices and quantum technologies \cite{ADVMATpal2022}.

Despite several theoretical studies on strained arsenene \cite{NSzhang2015, PRBkamal2015, AIPADVwang2016}, experimental investigations remain very limited. Arsenene layers have been extensively produced by different methods, such as mechanical exfoliation, liquid-phase exfoliation, and chemical vapor deposition (CVD) \cite{SMALLhu2022}, the main limitation referring to the strain application. Axial strain in 2DMs is usually applied via extrinsic methods, by using elastic substrates or controlling the 2DM thermal expansion \cite{ADVMATdai2019, LSApeng2020, JPCMrafael2015}. These approaches are typically limited to small strains $\sim 6\%$ \cite{ADVMATdai2019} and the relatively weak interactions between 2DMs and most of the substrates make the application of large extrinsic tensions very challenging. 

In this paper, to tackle the challenge of producing highly strained arsenene, molecular beam epitaxy (MBE) was employed to grow arsenic onto a silicon carbide (SiC) surface. Due to the high control of the molecular beam flux and the growth temperature, MBE allows the production of high-quality 2DMs where large intrinsic strains can be achieved via the lattice mismatch between the target 2DM and the selected substrate \cite{NATCOMMahn2017}. SiC was chosen as substrate due to its high chemical and thermal stability \cite{JJAPmatsunami2004, JACwei2018}, and crystal structure that allows lattice mismatch induced strain $>10\%$ with arsenene. We experimentally observed the formation of few-layer arsenene onto SiC, where the first monolayer covalently binds to the substrate, yielding an in-plane tensile strain of $23\%$. The combined experimental and density functional theory results allow the interpretation of a new arsenene triangular allotrope. The ab initio calculations show the energetically favorable formation of a highly strained arsenene first layer, depicting also a decoupled nature of further arsenene layer staking. The reported system put in evidence strategies towards the development of highly strained 2DMs.

\section{Methods}

The As deposition was carried out in a customized MBE chamber from Dr. Eberl MBE-Komponenten dedicated to growing III–V compounds. A commercial epi-polished 6H-SiC(0001) n-dopped wafer (miscut on-axis $\pm 0.5^o$), Si-terminated, was cleaved in pieces of $8\times8$\,mm and used as substrate. The SiC surface is characterized by $0.4$\,nm high Si-C bilayer terraces ($150$\,nm wide), exhibiting an overall $< 0.5$\,nm surface roughness (over $2''$ wafer area). The SiC substrate pieces were attached to a $2''$ Si wafer using eutectic gallium-indium liquid metal alloy on a holder facing down the As$_4$ flux. To remove residual impurities from the surface, the substrate was annealed for $1$\,h at $400^o$\,C and dry-etched in an ultra-high vacuum (UHV) using a hydrogen beam. The etching procedure used a thermal cracker to source an ion-free atomic hydrogen gas beam toward the substrate surface that was heated at $700^o$C and exposed to a hydrogen background pressure of $10^{-5}$\,mbar for $10$ min. Prior to the As epitaxial growth, the substrate was annealed at $850^o$C in the deposition chamber (base pressure $2 \times 10^{-10}$\,mbar) to remove the inert chemical H-termination and consequently uncover the SiC dangling bonds. The substrate was cooled down to $160^o$C and exposed to As$_4$ molecular beam flux. A low-temperature effusion cell was used to sublimate arsenic solid source ($99.999995\%$ purity metallic arsenic from Furukawa Denshi Co., LDT.) at $320^o$C to reach the desired beam equivalent pressure (BEP) measured by a hot‑filament ionization gauge placed in front of the beam.

A thin Al capping layer was deposited in situ immediately after the arsenene growth for the analysis of the sample cross-section via an external high-resolution transmission electron microscope (HRTEM). The HRTEM specimens were prepared by focused ion beam (FIB) and imaged using a Thermo Fisher FEI Titan HRTEM operating at $300$\,kV. Atomic force microscopy (AFM) characterization was carried out in a Bruker Multimode instrument in peak force tapping mode using a ScanAsyst-Air (Bruker) tip with $0.4$\,N/m spring constant operating at room temperature and in a nitrogen-rich atmosphere. Chemical ex-situ analyses were carried out by X-ray photoemission spectroscopy (XPS) at room temperature using a Thermo Scientific K-alpha spectrometer with Al K$\alpha$ micro-focused monochromator (energy $1486.6$\,eV). The X-ray spot size was settled at $400$\,{$\mu$}m. Wide-scan surveys of all elements and high-resolution spectra for As, Al, Si, C, and O were recorded at a base pressure of $10^{-7}$\,mbar. AFM and XPS analyses were carried out in arsenene samples absent of the Al capping layer. The oxide formed on arsenene was removed in the XPS chamber using Ar ion beam at $1$\,keV.


DFT calculation was performed in the plane wave basis Viena Ab initio Simulation Package (VASP) \cite{PRBkresse1993, PRBkresse1996}. All plane waves with energy below $400$\,eV were taken into consideration in the base. The electron-ion coupling is included through the projected augmented wave (PAW) method, where all atoms were allowed to relax their position until all forces become lower than $2.0$\,meV/{\AA}. The exchange-correlation term was described using the generalized gradient approximation (GGA-PBE) \cite{PRLperdew1996}. The total energy was sampled in a 2D Brillouin zone with a k-point density of 619\,{\AA}$^2$. We considered van der Waals interactions in all calculations within the nonlocal vdW-DF functional \cite{PRLdion2004}. The SiC substrate surface was described by a slab method considering six atomic layers \cite{CARBONpadilha2019}. Unless otherwise stated, all arsenene calculations was performed on top of a SiC slab. The periodic image interaction was avoided by using at least $20$\,{\AA} vaccum perpendicular to the surface.

\section{Results}

\begin{figure*}
    \includegraphics[width=1.5\columnwidth]{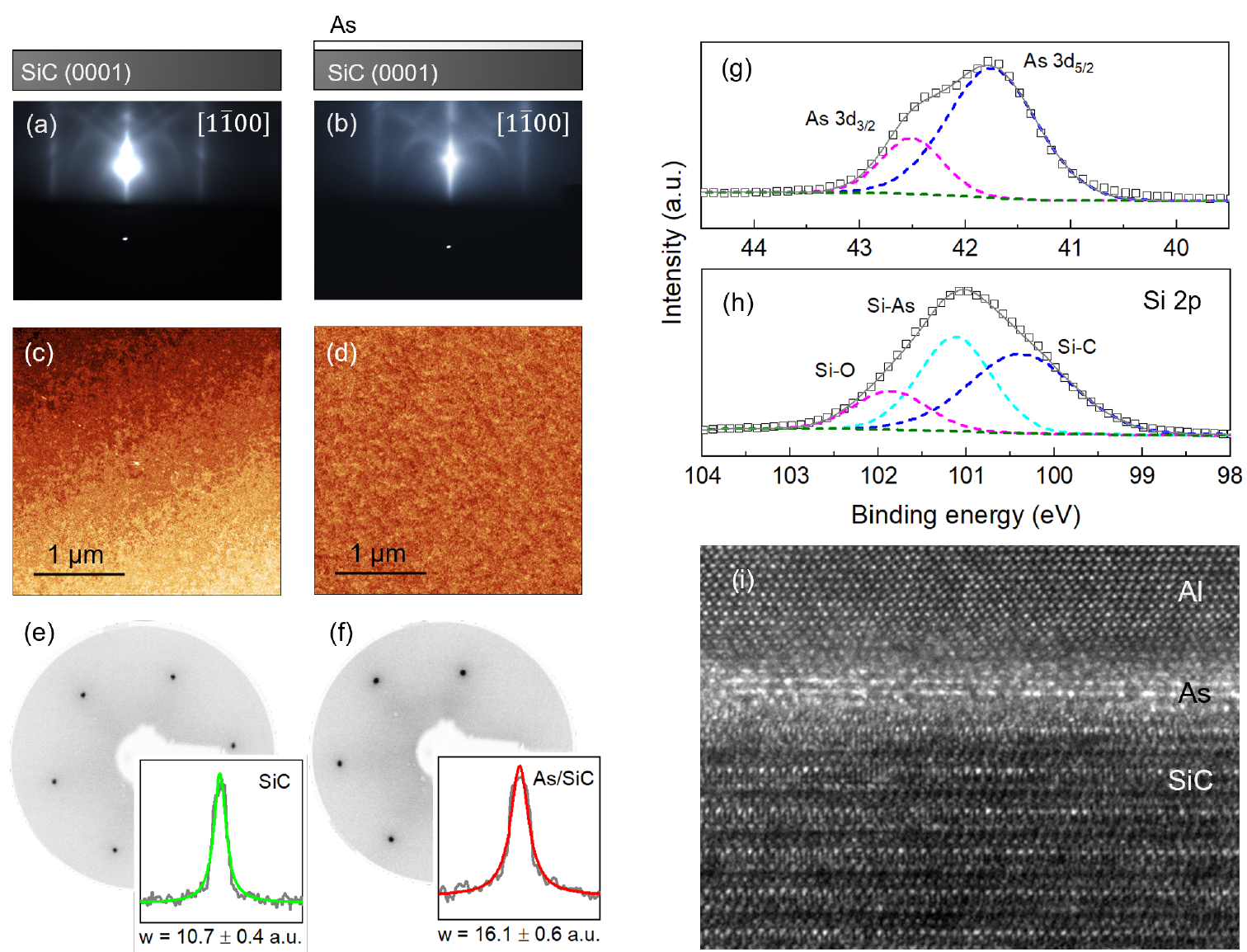}
    \caption{\label{fig1} (a) RHEED pattern of the bare SiC substrate after annealing at $850^o$C and (b) the As deposition the diffraction pattern. The illustration above (a) and (b) refer to the sample structure for each RHEED image. (c) AFM image of the bare SiC substrate revealing the Si-C bilayers surface steps with 0.4 nm of height. (d) AFM image after As growth. (e) $1\times1$ LEED diffraction pattern from 6H-SiC (0001) after annealing and prior to As growth, and (f) after the As deposition. LEED patterns were obtained at 90 eV. High-resolution XPS spectrum of (g) As 3d energy range and (h) Si 2p. (i) TEM cross-section image of the grown arsenene on SiC having an Al capping layer.}
\end{figure*}

Figure~\ref{fig1}(a) shows the RHEED pattern of bare SiC. The typical SiC surface step array is characterized by both the observed diffraction streaks and the splitting of the specular point. For substrate miscut angle lower than $10^o$, the beam splitting can be seen at any azimuthal direction but parallel to the terrace edge \cite{ASSjian1999}. In addition, Kikuchi lines from the SiC bulk are observed evidencing the high crystalline quality of the substrate. The presence of more than one specular spot due to the SiC terraces causes the central region in Fig.~\ref{fig1}(a) to be broad and highly bright. After the As deposition, the central spot becomes thinner [Fig.~\ref{fig1}(b)] indicating the vanishing of the SiC terraces. This result is corroborated by AFM images which exhibit clear terraces on the bare SiC [Fig.~\ref{fig1}(c)] and a stepless homogeneous surface after As deposition [Fig.~\ref{fig1}(d)]. The surface roughness of the deposited As shown in Fig.~\ref{fig1}(d) was found $0.84$\,nm rms over a $9$\,$\mu{\rm m}^2$ scanned area. No significant differences in the RHEED streak distances were observed during and after the As deposition, suggesting a layer-by-layer growth mechanism for the deposited arsenene. To allow layer-by-layer growth, the As bond with substrate should be stronger than the As-As interaction \cite{SCTkirill2020}, yielding arsenene to be formed on SiC. {Indeed, our ab initio calculations confirm an exothermic process of As$_4$ bonding to SiC surface.} Free-standing planar arsenene has been predicted to not be dynamically stable \cite{PRBkamal2015, SRrahman2019}. 

LEED patterns were also recorded prior to and after the As deposition. As shown in Fig.~\ref{fig1}(e), sharp LEED spots in a low diffuse background were observed for bare SiC. This pattern, from the unreconstructed $1\times1$ SiC after annealing at $850^o$C, indicates a high-quality (smooth and well-organized) substrate surface. After the As deposition, no reconstruction was observed [Fig.~\ref{fig1}(f)], corroborating that the arsenic layers grow with the same orientation of the SiC unit cell. The LEED spot profiles [inset in Fig.~\ref{fig1}(e) and (f)] show a significant broadening of the diffraction peak from $10.7 \pm 0.4$\,a.u. to $16.1 \pm 0.6$\,a.u. for the full width at half maximum (w) when As is deposited. Such an increase can be related to the overlap of the diffraction spots of the SiC substrate and that of the grown arsenene, indicating a slight increase of the surface disorder upon the As deposition. Finally, from the distances between $1\times1$ dots on the LEED pattern, we find that the lattice constant (a) of the deposited arsenene is $3.1$\,{\AA}. This value is very similar to that reported in the literature for bare hexagonal SiC ($a = 3.08$\,{\AA}) \cite{JCGjones2008}. This result clearly indicates that the epitaxially deposited arsenene follows the very same surface structure of the SiC. {Here it is worth noticing that arsenene is in contrast with the honeycomb reconstruction observed on SiC bismuthene, which can be understood by our ab initio results below.} From the lattice mismatch between 6H-SiC nearest neighbor sites (d$_{\rm NN} = 3.08$\,{\AA}) and that of freestanding two-dimensional buckled arsenene (d$_{\rm NN} = 2.50$\,{\AA}) \cite{PRBkamal2015, JCGjones2008} the induced tensile strain on the deposited arsenene can be estimated as $23\%$.

Highly strained arsenene could be epitaxially grown thanks to the strong As interaction with SiC. In order to examine such an interaction, XPS analyses have been carried out. Fig.~\ref{fig1}(g) and (h) show the respective high-resolution As 3d and Si 2p XPS spectra of the sample after etching at the XPS chamber. As 3d$_{3/2}$ and As 3d$_{5/2}$ peaks are observed at $42.5$ and $41.7$\,eV, respectively \cite{SSrochet1995}. The absence of a third peak at $44.9$\,eV typically ascribed to oxidized As (e.g. As$_2$O$_3$) \cite{APLMATchen2021, NSADVantonatos2020, NISTxps} demonstrate that the etching procedure efficiently removed the As oxide layer. Concerning the Si 2p XPS spectrum, the peaks identified correpond to SiC, SiAs and SiO, at $100.2$\,eV, $101.1$\,eV and $102.1$\,eV\cite{SSrochet1995}, respectively [Fig.~\ref{fig1}(h)]. The presence of SiO may refer to the remaning oxygen passivation layer onto SiC not completely removed during the hydrogen etching.  The identification of the Si-As peak is evidence of the covalent bonding formation between SiC and arsenene. The covalent bonding between Si and As has been corroborated by DFT calculations, as discussed hereafter. Finally, the epitaxial growth of arsenene on SiC has also been probed by HRTEM cross-section imaging as shown in Fig.~\ref{fig1}(i). The HRTEM image shows a few-layer arsenene sandwiched between the SiC substrate and the Al capping layer. From Fig.~\ref{fig1}(i), however, no distinction between the common arsenene allotropes (viz. planar, buckled, and puckered) is possible. The uppermost interface of the arsenene layers is not sharp likely due to the energetic Al atoms deposited during the capping layer preparation. Additionally, the arsenene/SiC interface is subject to the substrate roughness.

\begin{figure}
    \includegraphics[width=\columnwidth]{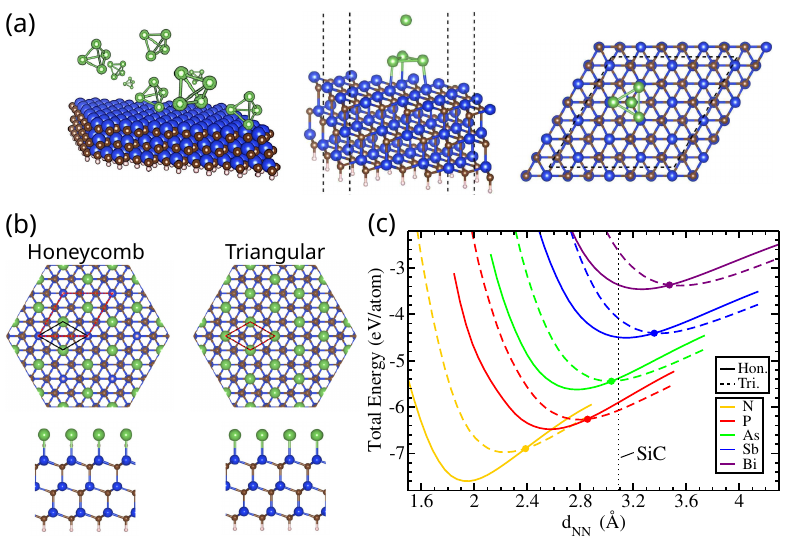}
    \caption{\label{fig:hex-tri} (a) As$_4$ configuration on SiC surface. (b) Possible periodic structures of As on SiC, forming a honeycomb geometry or a triangular lattice. (c) Relative energy between honeycomb (continuous line) and triangular (dashed lines) lattice for the pnictogen family.}
\end{figure}

Despite flat antimonene and bismuthene experimental growth on SiC surfaces their flat As counterparts were grown and theoretically explored on metallic surfaces \cite{2DMATshah2020, ASSkang2019}. Here, in our MBE experiments, a flux of As$_4$ molecules is directed to the SiC substrate. Our ab-initio calculations indicate an energetically favorable binding energy of the As$_4$ molecule on the SiC surface, with a binding energy of $-2.29$\,eV/As. In Fig.~\ref{fig:hex-tri}(a) we show the final configuration of As$_4$ on the SiC surface, indicating a possible triangular structure, as the As binds to the nearest-neighbor Si surface lone-pairs. Our experimental electron diffraction results indicate that the presence of the As on the SiC surface does not lead to reconstruction, that is, the As layer follows the bare SiC surface periodicity. A different scenario is described in the literature for the Sb and Bi cases, where a honeycomb lattice reconstruction is observed \cite{SCIENCEreis2017, NLshao2018}. We have calculated the formation energy ($E_{\rm for}$) of a honeycomb and a triangular phase of As on SiC [Fig.~\ref{fig:hex-tri}(b)]
\begin{equation}
    E_{\rm for} = \frac{1}{N} \left(E_{\rm final} - E_{\rm initial} - \frac{N}{4}E_{As_4}\right),\label{eq:for}
\end{equation}
where final/initial systems are the SiC slab with/without the arsenic first layer, $E_{{\rm As}_4}$ is the total energy of an isolated As$_4$ molecule, and $N$ the number of As per unit cell in the first layer. The triangular phase is energetically favored by $0.13$\,eV/As, with $E_{\rm for} = -1.59$ and $-1.72$\,eV/As for honeycomb and triangular, respectively. Here we can understand the deviating behavior of As compared to Sb and Bi structures by comparing the energetics of a honeycomb lattice and a triangular lattice for the pnictogen family as a function of the interatomic distance. Notice here that in order to capture the substrate effect we have replaced the SiC slab with passivation of the bottom part of each layer with hydrogen, which is shown to correctly capture the pnictogen/SiC systems properties \cite{NJPhsu2015, NLcosta2019}. In Fig.~\ref{fig:hex-tri}(c) the vertical dashed line marks the SiC surface lattice nearest-neighbor (NN) distance. For the Sb and Bi atoms, which present a larger atomic radius, the lower site density honeycomb lattice, is more stable on the SiC lattice parameter. However, for the lower atomic radius pnictogens, a lattice with a higher site density is more stable, stabilizing the triangular reconstruction. Interestingly such analysis is consistent with the observed honeycomb lattice of As on Cu, Ag, and Au (111) surfaces \cite{2DMATshah2020, ASSkang2019}, which presents d$_{NN}$ between $2.5$ and $2.9$\,{\AA}, that is, in the region where the honeycomb lattice is energetically favorable, Fig.~\ref{fig:hex-tri}(c). 

\begin{figure}
    \includegraphics[width=\columnwidth]{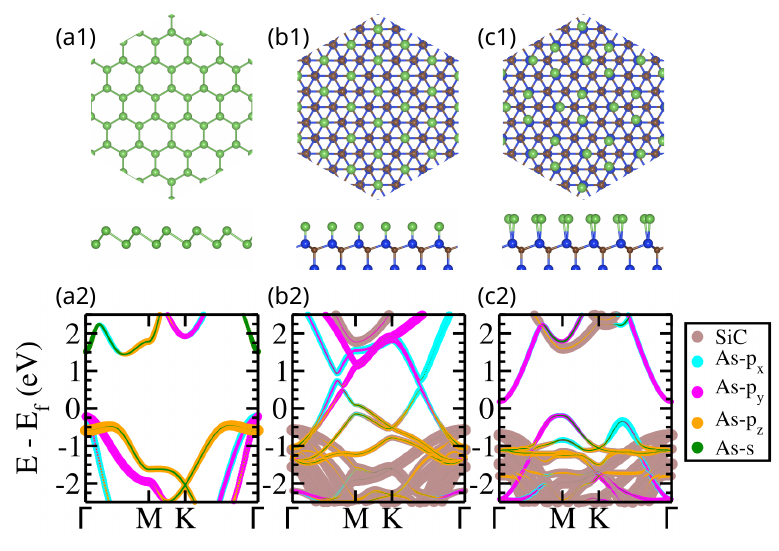}
    \caption{\label{fig:bnd-str} Atomic structure and orbital projected band structure for freestanding arsenene (a1)/(a2) and triangular phase over SiC symmetric (b1)/(b2) and trimerized (c1)/(c2).}
\end{figure}

From the electronic perspective, the freestanding arsenene in its buckled structure [Fig.~\ref{fig:bnd-str}(a1)] is known for being a topologically trivial semiconductor, Fig.~\ref{fig:bnd-str}(a2). Previous studies have discussed that tensile strain over $13\%$ in arsenene a topological trivial $\rightarrow$ non-trivial transition would be present \cite{SRrahman2019}. However, as discussed the arsenene system goes through a structural phase transition [honeycomb $\rightarrow$ triangular] in $12\%$ strain, Fig.~\ref{fig:hex-tri}(c). Here, the arsenene bands neatly lie within the SiC band gap, Fig.~\ref{fig:bnd-str}(b2) and (c2). The symmetric triangular structure presents a metallic phase with a vanish velocity band (high density of states) resonant with the Fermi energy along the $M-K$ dispersion, Fig.~\ref{fig:bnd-str}(b2). This electronic picture gives rise to a Peierls-like structural transition where a gap of $0.37$\,eV is opened on the As-p states, Fig.~\ref{fig:bnd-str}(c2). Such Peierls-like trimerization presents an atomic distortion with average atomic dislocation of $0.2$\,{\AA}, and an energetic preference of $-0.13$\,eV/As [Fig.~\ref{fig:bnd-str}(c1)].

\begin{table}
\begin{ruledtabular}
    \caption{\label{tab:2L} Second layer formation energy compared with the As$_4$ molecule ($E_{\rm for}^{(2L)}$) \ref{eq:for}. Second layer interaction energy with the first one $E_{int}$, where negative/positive energies dictate an attractive/repulsive interaction. The interaction energy is defined as the formation energy minus the strain energy of the second layer compared with a freestanding system.}
    \begin{tabular}{lcc}
         interface & $E_{for}^{(2L)}$ (eV/As) & $E_{\rm int}$ (meV/As)  \\
         \hline
         $T(2)/H(\sqrt{3})$          & 1.99 & -32\\
         $T(2)/H(\sqrt{7})$          & 0.05 & -9 \\
         $T(3)/H(2\sqrt{3})$         & 0.36 & -20 \\
         $T(3)/H(3)$                 & 1.00 & -45 \\
         $T(\sqrt{7})/H(3)$          & 0.36 & -22 \\
         $T(\sqrt{7})/H(4)$          &-0.22 & -52 \\
         $T(4)/H(2\sqrt{7})$         & 0.06 & 0 \\
         \hline
         $T(2)/T(2)$                 & 0.80 & -59 \\
         $T(\sqrt{7})/T(\sqrt{7})$   & 0.78 & -59 \\
         $T(2)/T(\sqrt{7})$          & 1.73 & 1 \\         
    \end{tabular}
\end{ruledtabular}
\end{table}

The experimental results are also suggesting that the system does not grow thicker arsenene slabs. We have explored the multilayer arsenene system supported by SiC. Note that the first arsenene layer is bonded covalently with the SiC substrate in a triangular structure with exfoliation energy of $292.5$\,meV/{\AA}$^2$. In order to understand the energetically stable second layer we have explored different cell mismatches both in a Triangular/Triangular interface as well as a Triangular/Hexagonal interface, which we summarize in Table~\ref{tab:2L}, where the first Triangular layer is constrained with the SiC lattice periodicity. We have found that a second layer hexagonal $4 \times 4$ cell [$H(4)$] matched with an triangular $\sqrt{7} \times \sqrt{7}$ cell [$T(\sqrt{7})$] interface, namely $T(\sqrt{7})/H(4)$, is the most stable system explored. We can understand such preference arising both from the lower strain in the second layer [$<2\%$], as well as an increased interaction energy $-52$\,meV/As. The interaction energy is inferred from the second layer formation energy, ruled by interaction plus strain energy, $E^{(2L)}_{\rm for} = E_{\rm strain} + E_{int}$, being the interaction energy always negative, attractive vdW coupling, while the strain energy is always positive. Although in the smaller commensurable system $T(2)/T(2)$ and $T(\sqrt{7})/T(\sqrt{7})$ a stronger binding energy is achieved, the lower strain presented on the $T(\sqrt{7})/H(4)$ dictates the favorable formation energy. From the results, we can understand that a perfectly decoupled second layer is the more stable configuration.

\begin{figure}
    \centering
    \includegraphics[width=\columnwidth]{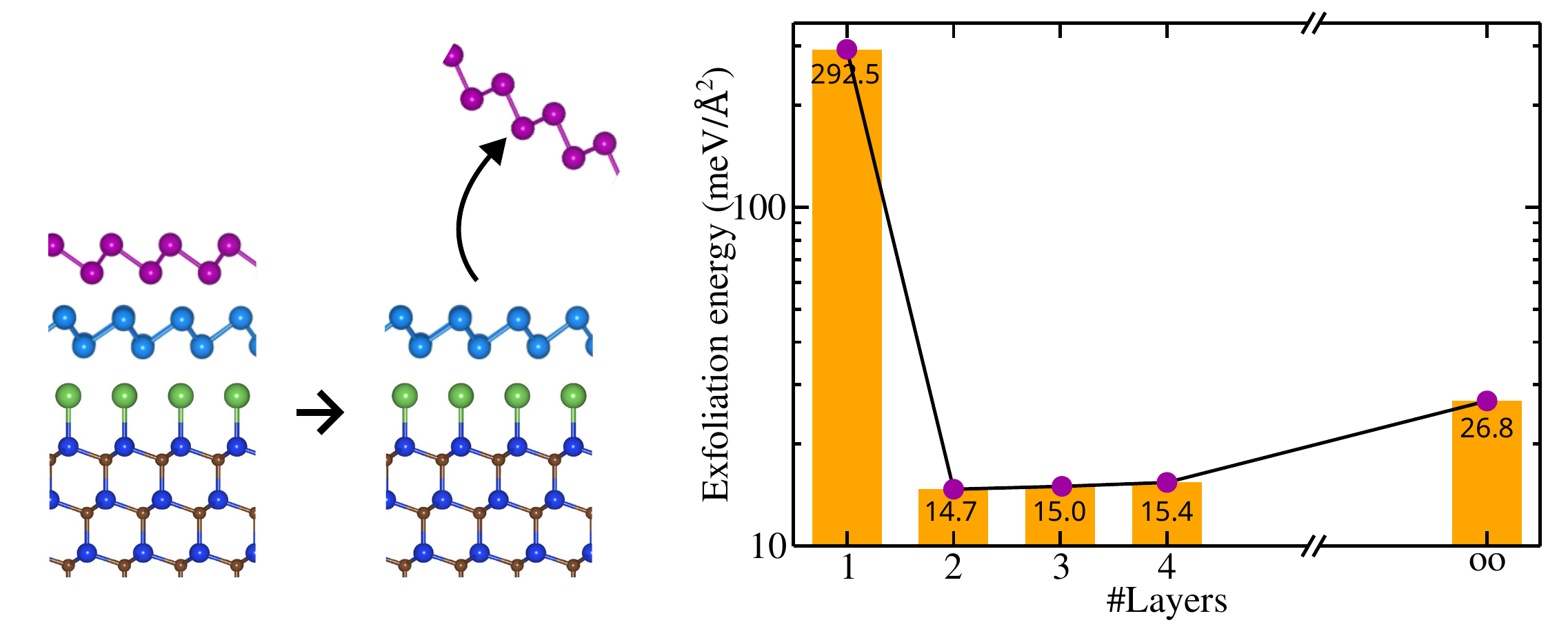}
    \caption{\label{fig:exf-ene} Exfoliation energy for removing one arsenic layer over N-layer slab. The first layer is covalently bonded to SiC.}
\end{figure}

The decoupled character of the second layer present a binding energy ruled by vdW interaction giving a exfoliation energy of only $14.7$\,meV/{\AA}$^2$. Such scale of values is maintained for 3 and 4 stacked layers, with the exfoliation energy of the top-most layer being, for instance, lower than that calculated for graphene $20$\,meV/{\AA}$^2$. In contrast, the gray arsenic bulk presents a exfoliation energy of $26.8$\,meV/{\AA}$^2$ [Fig.~\ref{fig:exf-ene}]. Here we can expect that, despite gray arsenic being stable, few layers arsenene presents a lower interlayer interaction leading to a possible self limited growth.

\section{Conclusions}

In summary, we reported the experimental synthesis of 2D arsenene supported by SiC. Our electron diffraction results predict a new triangular arsenene allotrope which was confirmed by our ab initio calculations.  The strain field arising in the arsenene first layer over silicon carbide drives a structure transition from the honeycomb freestanding system to the supported triangular. The silicon carbide, beyond serving as a generator of the strain field, with its insulating electronic character allows for the relevant arsenene state to neatly lie within its energy gap. Such a new triangular allotrope, present a Peierls-like distortion, leading to an indirect semiconducting character with a $0.37$\,eV. Our ab initio results also indicate a possible self-limited epitaxial growth of a few layers arsenene given its decoupling of the second non-covalent layer.

\begin{acknowledgments}

The authors acknowledge financial support from the S\~ao Paulo Research Foundation FAPESP (Grant numbers 21/06238-5, 19/14949-9, 19/20857-0, and 17/02317-2), INCT-Nanocarbono, INCT-Materials Informatics, and Laborat\'{o}rio Nacional de Computa\c{c}\~{a}o Cient\'{i}fica for computer time (project ScafMat2). R.F.O. acknowledges additional support from INCT/INEO. 

\end{acknowledgments}


\bibliography{bib}

\end{document}